\documentstyle[manuscript,aps]{revtex}
\begin{document}

\title{Enhancement of parity and time invariance  violation in Radium atom}

\author{V.V. Flambaum \thanks{email address: flambaum@newt.phys.unsw.edu.au}}

\address{ School of Physics, University of New South Wales,
Sydney 2052, Australia}

\maketitle

\tightenlines


\begin{abstract}
There are several factors which lead to a huge enhancement 
of parity and time invariance violating effects in the Ra atom:
very close electronic levels of opposite parity, 
the large nuclear charge Z and the collective nature of T,P-odd
nuclear moments. Experiments with Radium may be used
to  measure  it's nuclear anapole, magnetic
quadrupole and Schiff moments. Such measurements
provide information about parity and time invariance
violating nuclear forces and electron-nucleon interactions. 
\end{abstract}



  Effects of time invariance (T) and parity (P) violation are inversely
proportional to the distance between  opposite parity energy
levels. There is a pair of close  opposite parity levels in the Ra atom,
the state $|1> =|7s6d$ $J^P=2^+>$, with  E=13993.97
$cm^{-1}$ and the state $|2>= |7s7p$ $J^P=1^->$, with  E=13999.38 $cm^{-1}$,
  which are separated by a very small interval
(5 $cm^{-1} \sim 10^{-3}$ eV ). Note that there are also very close levels
 of opposite parity in rare-earth atoms (e. g. in Dy atom \cite{Dzuba,Budker})
which have been used to measure parity violation.
 The advantage of Ra 
is that the electron states here are simple. Therefore, the mixing of these
 states by
the weak interactions 
is not suppressed and can be accurately calculated
 (contrary to the mixing in the rare-earth atoms
where there is a strong suppression of the matrix elements due to the extreme
complexity of the electronic eigenstates). There are also extra factors of
enhancement in Ra: the large value of the nuclear charge, Z=88 (the P and T
odd effects increase with Z faster than $Z^2$) and the collective 
T,P-odd  moments of deformed Ra nuclei that are much greater than the same
 moments in spherical nuclei \cite{Flambaum,Auerbach}.
This  makes Radium an attractive object for future experiments.

  The close electron levels in Ra have different electron angular momenta,
J=1 and J=2. The  conservation of the total angular momentum ${\bf F} = {\bf
J} +{\bf I}$
requires the involvement of the nuclear spin $I$ for these states to be mixed.
Parity violating effects in this case can be produced by the nuclear
anapole moment, which is directed along the nuclear spin.
An atomic electric dipole
moment ( EDM) appears due to the interaction between the atomic electrons and 
nuclear T,P-odd moments: magnetic quadrupole, Schiff 
and electric octupole moments. These effects also can appear due to P,
T-odd nuclear-spin-dependent electron-nucleus interactions.

 Let us start from an estimate of the
contribution of the nuclear magnetic quadrupole moment $M$. Atomic EDM
in the metastable state $|1>$ appears due to mixing with the opposite parity
 state $|2>$
by the  Hamiltonian $H_M$ of the magnetic interaction between the nuclear
magnetic quadrupole moment and the atomic electrons. The atomic EDM is given by
the following formula   
\begin{equation}
\label{EDM}
{\bf d}=2 \frac{<1|H_M|2><2|-e{\bf r}|1>}{E_1 -E_2}
\end{equation}
The spin-orbit interaction in Radium is very large, therefore the electronic
states can be approximately described by  jj-coupling. In this
case $|1> \simeq |7s6d_{3/2}, J=2>$ and $|2> \simeq |7s7p_{1/2}, J=1>$.
An expression for the matrix elements of $H_M$ between the electronic
orbitals  $6d_{3/2}$ and $7p_{1/2}$ can be found in \cite{FKS84}.
Using this expression and simple numerical estimate for the matrix element
of the radius vector ${\bf r}$ between these orbitals we obtain
the following value for the Ra atomic EDM:
\begin{equation}
\label{EDMM}
d \simeq 0.5 \cdot 10^3 M m_e
\end{equation}
where $m_e$ is the electron mass.The magnetic quadrupole moment $M$
is not zero in those Ra isotopes that have 
 nuclear spin $I \geq 1$, for example,
in $^{223}Ra$ where $I=3/2$. This nucleus is deformed, therefore, it has a
collective magnetic quadrupole moment $M$, which was
 estimated in Ref. \cite{Flambaum} to be
$M \simeq 10^{-19} cm \cdot \eta \, e/m_p$, where $m_p$ is the proton mass and
 $\eta$
is the dimensionless constant of the T,P-odd nucleon-nucleon interaction
(the strength of the T,P-odd interaction is  measured in units of the weak
 interaction Fermi constant). The final result for the atomic EDM produced
by the T,P-odd nucleon-nucleon interaction is
\begin{equation}
\label{EDMeta}
d \simeq 2 \cdot 10^{-20} \eta\, e\, cm
\end{equation}
This value is about $3\cdot 10^4$ times larger than EDM of $Hg$ atom,  which
was measured in Ref. \cite{Hg} and gives the present best 
 limit on $\eta$.     

 The  atomic EDM can also be produced by the nuclear Schiff moment (see, e. g.
 \cite{FKS84}). However, the Schiff moment produces a contact interaction
only (an electric field inside the nucleus). The mixing of
 $6d_{3/2}$ and $7p_{1/2}$ states by this interaction is very small.
However, one can take into account configuration mixing and use the maximal 
 matrix element between the $7s$ and $7p_{1/2}$ states (for example,
include mixing between the $7s6d$ and $7p7p$ configurations).
Also, there is theoretical and experimental evidence that the
 odd Radium isotopes  have octupole deformation, which leads to
a huge $10^3$ enhancement of the nuclear Schiff moment \cite{Auerbach}.
A rough estimate of the Schiff moment contribution that includes this 
enhancement gives a value of the atomic EDM comparable to the
 magnetic quadrupole contribution (\ref{EDMeta}). The advantage of
 the Schiff moment  is that 
it exists also for the isotopes with nuclear spin I=1/2, like $^{225}Ra$
 (the magnetic quadrupole is equal to zero for such isotopes). 

  Finally, Radium atom can be used to measure the nuclear anapole moment.
In principle, there are several possibilities. One can measure,
for example, interference between the Stark and parity violating amplitudes
 in the transition between the ground
 state $|0>=|7s^2, J=0>$ and
excited state $|1> =|7s6d$ $J^P=2^+>$,  E=13993.97
$cm^{-1}$.
 The parity violating amplitude
 $E1_{pv}$ appears due to the weak mixing
between the state $|1>$ and the opposite parity state $|2>= |7s7p$ $J^P=1^->$,
  E=13999.38 $cm^{-1}$.
  Again, one has to take into account the configuration mixing since
the direct matrix element $<6d_{3/2}|H_a|7p_{1/2}>$ of the  interaction
$H_a$ between the atomic electrons and the anapole magnetic field
which is localized inside the nucleus, is very small. A rough numerical
estimate shows that due to the closeness of the opposite parity levels 
and the large nuclear charge Z the amplitude $E1_{pv}$ in Ra is
several hundred times larger than similar
amplitude  ( induced by the anapole) in  Cs; the latter
 amplitude was recently measured in
 Ref. \cite{Wieman}.

Note that in Ra the anapole-induced amplitude gives the dominating
 contribution  to the total 
parity violating amplitude.
This may be an advantage since in Cs the anapole contribution had to be
 separated
from the weak charge contribution  ( due to Z-boson exchange
between electrons and nucleus) which was two orders of magnitude larger
than the anapole contrubution. We may neglect the weak charge contribution
in Ra since the states $|1>$ and $|2>$  have different electron angular
 momenta J=2 and J=1 and cannot be mixed by the weak
 nuclear-spin-independent interaction.
 
 In conclusion, we considered strongly enhanced parity and time invariance
violating effects in Ra atom. Unfortunately, all odd Radium isotopes
are unstable. However, recent progress in trapping of unstable
elements  makes such experiments feasible. For example,
an experiment on parity violation in the short-lived Fr atom is in progress
 \cite{Fr}. An experiment with $^{225}Ra$, motivated by the work
 \cite{Auerbach},
 has been discussed  by S. Lamoreaux \cite{Ra} and A. Young \cite{Young}.
The Los Alamos laboratory
has a source of this isotope. Note also that the electronic structure
of Radium atom is relatively simple (two electrons above
closed shells).  Our experience has shown that
the accuracy of atomic calculations for such systems can be
about one per cent (see, e.g. \cite{DFK}).  

This work was supported by the Australian Research Council.

\end{document}